# A Case-Based Look at Integrating Social Context into Software Quality

Nicole Radziwill, Morgan Benton, Kenneth Boadu, and Wilson Perdomo

*Ensuring high-quality software requires considering the social climate within which the applications will be deployed and used. This can be done by designing quality goals and objectives that are consistent with changing social and ethical landscapes. Using principles of technological determinism, this article presents three cases that illustrate why it is becoming even more important to integrate these concerns into software design and quality assurance. With these examples in mind, this article explains how to consider technological determinism in software design and quality assurance practices to achieve this enhanced sensitivity on a practical level.*

## INTRODUCTION

According to Winner (1980), "Technical things have political qualities" and software is no exception. Perhaps the clearest evidence of this can be seen from the "Arab Spring" uprisings across the Middle East mediated by social media (Stepanova 2011), or in the many protests across China that were organized informally by Twitter users (Liu 2014). Software designers have the power to shape millions of lives, and with this comes great responsibility: "Through software, it has become possible for a non-political elite, upholding a non-political ideology, to dominate the whole society…to control the means to represent the world [and how] we acquire knowledge, communicate, and conduct our affairs" (Sorin 2013).

In the process of envisioning and designing the software that drives many of today's transformational technologies, Winner argues that people often overlook the more long-term human consequences of development and deployment. In particular, most people do not reflect on how these can influence power dynamics, wealth distribution, social mobility, gender equality, and race relationships. He goes on to explain that technologies provide structures for human activities to achieve desired outcomes, and as a result, *technological determinism* is enacted: the technologies people implement actually determine how they will act and interact, and serve as the basis for forming culture. Leika et al. (2013) explain that these decision points arise because "technologies – when understood as combinations of human action and the technical artifacts or systems in use – are always designed for people, to be used for improving the quality of their everyday life."

This article presents three cases to underscore why it is becoming more important to integrate human and social concerns into software design and quality assurance. In the first case, the rapid spread of mobile technology throughout Africa shows how introducing new technologies can change the nature of social interactions and power dynamics. Next, intelligent vehicles (and self-driving cars) are used to illustrate how unexpected social issues can emerge from technology adoption. Finally, autonomous weapons show how technological advances can change the roles of, and relationships between, humans and machines. Each of these factors points to the need to more consciously address human and social issues as software is being designed, implemented, and deployed, because software is at the heart of each of these advances.

## APPLICATION DEVELOPMENT IN THE EMERGING AFRICAN MARKET

New technologies can change the nature of social interactions and power dynamics. The influx of technologies in general into the countries of Africa has caused a distinct social evolution, but the impact of mobile phone introduction has been unparalleled. While other Western technologies such as genetically engineered seeds, solar cook stoves, and innovative agricultural techniques have been implemented and accepted with varying success, mobile phones have been adopted quickly and (nearly) universally (Aker and Mbiti 2010). As a result, the ubiquity of cellular networks is dramatically reshaping the social world of the residents of Africa, in both positive and negative ways. The improvements and benefits are typically economic; the disadvantages are related to the rapid change of each culture's traditional methods of interaction and communication.

**Positive Impacts**
The chief and most obvious benefit of mobile devices for Africans is improved communication. It is important to note that African residents are not using cell phones simply for conversation, but have purchased one or more devices for use in day-to-day activities. According to the Afrobarometer, an independent, nonpartisan, African-led research project that measures the social, political, and economic atmosphere in Africa based on public opinion surveys, 77 percent of respondents said they owned their own mobile phone in 2013 (Bratton 2013), and the market was rapidly growing. Analysts expect that within a decade, the market penetration in Africa could exceed 90 percent, approaching the levels observed for Americans (Madrigal 2013).

The cell phone has quickly gained a permanent place in the social and economic life of Africa, embedding into work life in ways that communications technologies from the past several decades have not been able. Cell phones are most often used throughout different areas of the work force. Farmers and small business owners use their devices to contact clients and suppliers from distances they used to have to travel regularly (Donner 2006). Day laborers also experience the luxury of checking for work in cities before paying to make the transit there, formerly a prohibitive barrier to productive employment (Aker and Mbiti 2010). Hospitals and doctors allow patients to text in the event of emergencies or to receive daily reminders for taking medication.

Beyond the socio-economic impacts, there are additional interaction-based benefits of cell phone use. Families can now keep in contact from miles apart, and even communicate across national borders. Such a freedom in communication has never been seen before on the African continent, and could be compared to the rapid and pervasive adoption of computers in the Western world. In contrast, very few Africans have ever used a computer, and fewer have the ability to purchase and understand how to use one. The great benefit of cell phones is they allow Africans to use their own language in full context and emotion. Joint calls are extremely popular among family, but simple text messages also allow people to make connections. Cell phones are shrinking the world for the Africans who use them and making it possible to preserve relationships.

**Negative Impacts**
Mobile networks do not come without some detriment to the region. Growth and ease of communication is not always used for innocent and productive reasons. The use of cell phones among militant groups throughout Africa has accompanied a rise in violence over the last decade. Cell phones are an easy and unregulated way for armed factions to coordinate and execute plans of attack. It is unknown whether this is just a phase from the introduction of a new technology, or a more lingering problem. A similar trend has been observed in the Middle East.

Cell phones are also negatively affecting some social and cultural aspects of Africans' lives. While mobile devices are flooding into the region, not everyone can afford to own them or use them regularly. The possible result is for an opportunity gap to develop within African society where those with cell phones are able to thrive economically and socially, while those without suffer (Schlozman, Verba, and Brady 2010). With mobile networks becoming increasingly integrated into the African infrastructure this fear is quickly becoming a large threat to the social hierarchy of Africa, a change that could potentially destabilize local, regional, and national governance.

Mobile phones have also introduced a new social tension that has been unknown to the continent until this point. In a society so dependent upon relationships and social capital, cell phones have added new opportunities for communication and socialization. While this can be beneficial, it has increased general levels of distrust among married couples through "spouse spying" (Kenaw 2012). Husbands and wives can check each other's phones for signs of illicit behavior, or simply see who they have been contacting. Out of 50 Ethiopian couples interviewed by

Kenaw, only six said they did not check their spouses' devices, meaning that about 88 percent do. However, these individuals simultaneously reported that mobile devices had a positive impact on their relationships due to the ease of communicating. This highlights a significant cultural difference between Ethiopia and the United States where reports show that only about 25 percent of people admit to checking their partner's cell phone (the article discussing the U.S. phenomenon expressed the opinion that this was a high number), indicating a big difference in beliefs about privacy (Smith 2014).

In a society where women are given restricted freedom, a cell phone is another method for them to be subjected to the will of their husbands. It is common for men to dictate their wives' cell phone use, if they are even allowed to use one. On the other hand, mobile phones have allowed women to regulate the actions of their husbands in ways not possible before cellular networks. Wives tend to call their husbands daily to check up on things, and if the man does not answer when she knows he is free, then suspicions deepen. This allows spouses to have a social record of their partner's daily activities, which can lead to more anger. The articles cited in the previous paragraph suggest that this might be a more prevalent issue in African countries than in the United States.

**INTELLIGENT VEHICLES**
Unexpected *new* social issues can arise as well when new technologies are increasingly adopted, especially when the process involves innovation that transforms the experience of interacting with a product, object, or person. With that in mind, it is easy to overlook the several decades of innovation that have made intelligent vehicles part of life today. Adaptive cruise control, automatic climate control, anti-lock braking systems, automated fuel economy optimization, and "autopilot" for aircraft are all early examples of how advancements in intelligent vehicle research have already become mainstream. Although many researchers simply aim to augment the driver's ability to safely and effectively accomplish the driving task, some use techniques to enhance situational awareness that are as extreme as adding a robotic "co-driver" (Da Lio et al. 2015).

Like the market penetration of cell phones in Africa, self-driving cars can be expected to affect the majority of the U.S. population within the next two decades. In California, it is now legal for self-driving cars to share roadways with human drivers, and more states are expected to quickly follow suit; some analysts predict that self-driving cars will be out of the research stage and on the roads in an operational capacity by 2017 (Kelly 2012). Since self-driving cars use and extend the less controversial technologies that have already been introduced into intelligent vehicles, they provide an appropriate class of intelligent vehicles for impact analysis. This section examines the impacts of self-driving cars in the context of emergent social issues, and asks how software quality professionals can interject their skills and awareness to catalyze a more robust innovation process.

**Positive Impacts**
The more self-driving cars that are on the road, the more engineers and urban planners can take advantage of network effects. Traffic flow can be designed and optimized in ways that are not possible when it is necessary to factor in the uncertainty associated with human responses to traffic situations. Given that more than 1.2 million people die each year in traffic accidents, "Robot cars could be the cure for one of the world's biggest causes of deaths" (Dowling 2014). Self-driving cars can help to reduce the total number of car accidents that occur each year by eliminating errors due to driver distraction and fatigue.

They may also make long-distance driving less stressful and more enjoyable for all travelers. Hours of lost productivity previously spent driving will suddenly be available for other tasks. Environmental issues due to emissions will likely be drastically reduced. Industries will spring up to take advantage of newly realized possibilities that one can't even envision today (Kanter 2015).

Self-driving cars are also positioned to disrupt and potentially revolutionize the $200 billion auto insurance industry. If the potential for human error is eliminated, while volumes of data are automatically collected by the vehicle to resolve questionable cases, then liability for accidents (or, at the least, the "burden of proof") shifts completely to the manufacturer. Although this can reduce consumer costs and increase operational efficiency, it raises the possibility of new outcomes such as "distributed liability," which do not currently have precedents as models (Guerrini 2015). Along with this transfer of responsibility comes an increased burden on the designers and testers of the decision-making software embedded within.

Just because a technology makes things *easier* does not mean that the technology naturally supports social values that cultures wish to uphold. But once self-driving cars have a proven reputation for safety and reliability, people may come to see human drivers as a liability rather than a comfort.

**Negative Impacts**
While self-driving cars are beneficial in many ways, they also raise cause for concern. As robotics and software continue to advance, human labor can more readily be replaced by machines, eliminating jobs for bus, taxi, and delivery drivers. Kanter (2015) predicts that "Autonomous cars will be commonplace by 2025 and have a near monopoly by 2030, and the sweeping change they bring will eclipse every other innovation our society has experienced." This disruption is bound to stimulate new social issues as people who work as drivers migrate to the countries and regions that still operate traditional "manual" vehicles, causing potential economic and political upheaval as well. During the period of transition, however, it is likely that companies will still employ human agents to take over in the event of an emergency.

Most significantly, the widespread adoption of self-driving cars will shift a person's perception about the nature of control. People won't be able to travel anywhere without the car knowing who they are, where they went, and when they were in transit. Guerrini (2015) notes that driving one's own car could even become illegal, restricting the freedom of movement many people have come to expect. The sum total of this discomfort suggests that "the world is not quite ready to accept them – due to issues over liability, regulation or just the fact that people aren't yet ready to accept a car driving *them* to their destination… [and] police organizations and traffic safety experts are correct in saying there are risks in learning" (Dowling 2014).

The "risks in learning" cannot be underestimated. In fact, one of the most profound implications for this case is that, with intelligent systems, the testing and quality assurance process *does not end* even when all test cases have passed. Particularly with technologies that integrate artificial intelligence or machine learning, their full capabilities cannot be demonstrated until they have been "road tested" for quite a while and given the opportunity to adapt to their environments.

The most nefarious potential impact of self-driving cars, however, will be the threats to safety and security that can occur when the car is "hacked" by an outside source. In addition to prank-type hacks such as manipulating the air conditioning or turning the windshield wipers on and off, in some models, the car's engine and transmission can also be remotely manipulated with the right access. Although this could make remote diagnostics and repair possible, it also has the potential to threaten safety and wreak havoc on new models for determining liability (Greenberg 2015).

The role of the software developer or tester can evolve to accommodate these emergent social issues by considering the "hackers" to be stakeholders: both the ethical "white hat" hacker who aims to uncover and preemptively repair security issues, and the malicious "black hat" hacker whose purpose is to cause social, economic, and political harm.

**AUTONOMOUS WEAPONS AND KILLER ROBOTS**

Technological advances can also change the roles of, and relationships between, humans and machines. Although newer and not as well developed as the other cases, awareness of autonomous weaponry is starting to spread. In July 2015, a petition was spearheaded by physicist Stephen Hawking, entrepreneur Elon Musk, innovator Steve Wozniak from Apple, and Google DeepMind chief executive Demis Hassabis calling for a ban on development in this domain. The letter attached to the petition says that "AI technology has reached a point where the deployment of [autonomous weapons] is – practically if not legally – feasible within years, not decades, and the stakes are high: autonomous weapons have been described as the third revolution in warfare, after gunpowder and nuclear arms" (Gibbs 2015).

Some positive benefits are acknowledged. For example, "killer robots" may prevent battlefield deaths by fully taking humans out of the mission environment (similar to how drone pilots can now drop bombs in the Middle East from their computer screens on domestic military bases). But if software selects and destroys a real physical target without direct human intervention, who is morally and economically responsible for the damage? Similar to the case of self-driving cars, the implementation of autonomous weapons could lead to the emergence of new and unexpected social and political issues, but with far-reaching and potentially devastating consequences.

The pursuit of autonomous weaponry could change the perception of the nature of value, and more tightly connect people with the notion that value delivered at the individual, group, or nation level may not be aligned with value delivered to the human race or society as a whole (Holbrook 1996). It brings people back to the pre-World War II era to imagine how they might have responded then. If one were a developer or tester working on technology that led to the atomic bomb -- knowing the widespread death and destruction that came about as a result -- what might he or she have done differently? Over the next decade, many people working in the software industry may be faced with a similar question. "At such times it is the faith that 'man controls technology,' rather than the contrary view, which looms as an irrational belief" (Winner 1977).

**APPLICATION TO SOFTWARE DESIGN AND QUALITY ASSURANCE PRACTICE** The question remains: How can one more consciously address human and social issues in the software design and quality assurance processes for cases such as these? Following Winner's philosophy of technological determinism, these practices will be less useful in the future unless key questions are posed and answered, including:

- Will people be free to choose whether they want to use this technology? How can companies design and test software to preserve human involvement and interaction wherever possible?
- Does this technology enhance, rather than hinder, social values? Can people develop test cases that consider the *intent* of various user groups?
- Who is in control? How can people test to ensure that who they *intend* to be in control actually remains in control? If they relinquish control, who is liable?
- To what extent should this technology be expected to replace humans? Can developers design the system to help balance the involvement of human and machine in a way that preserves the value system?
- How can one reduce the risk that human invention will cause harm? Are there any risks due to early or incomplete adoption, which indicate that one should consider a more adaptive release?
- If hackers are stakeholders, what features should developers design into the system to help them accomplish their goals? How can companies integrate hacking behavior into the quality assurance process?
- How long is the learning process for this technology, and have developers designed test cases to accommodate that process after the product has been released? The testing process cannot find closure until the end of the "burn-in" period when systems have fully learned about their surroundings.

This list, however, is not exhaustive, meaning that integrating social context into practical concerns for software quality assurance is a ripe area for new research. "Despite the recent shift of focus to user experience," Leika et al.

(2013) explain that most contemporary human-technology interface research "concentrates still very much on usage situations." Their approach, motivated by Winner and other philosophers of technology who explored technological determinism, leverages design thinking to shift the emphasis from competence of use to stimulating quality of life.

According to these authors, design is powerful in that it encourages people to think about how life should be rather than in scientific approaches that focus on life as it is. This requires a value judgment that forces one to examine his or her relationship with a specific technology. Their four-pronged "life-based design" approach naturally integrates traditional software development and quality assurance activities into a wider framework that includes ethical analysis and quality of life considerations, and could be used as a basis for developing more contextually driven methodologies for software development and testing.

Finally, with the expectation that artificial intelligence and machine learning will become a more ubiquitous component of future applications, software development and quality assurance teams may need to expand their knowledge of the development and testing approach in mission-critical environments where safety, security, and regulatory concerns are paramount. This could involve re-examining attitudes toward agile development (ASQ 2011).

**Acknowledgments**
This article includes ideas and contributions from several students at James Madison University (JMU) in Harrisonburg, Virginia, who were enrolled in Morgan Benton's CS 330 (Computers in Society) course during the spring of 2015. They include Yvette Agyei, Angela Huynh, Hunter Taylor, and Michael Wood.